# Steady-State Probe-Partitioning FRET: A Simple and Robust Tool for the Study of Membrane Phase Behavior

Jeffrey T. Buboltz[1]

Department of Physics and Astronomy, Colgate University, Hamilton, New York, 13346

*Abstract*—An experimental strategy has been developed specifically for the study of composition-dependent phase behavior in multi-component artificial membranes. The strategy is based on steady-state measurements of fluorescence resonance energy transfer between freely diffusing membrane probe populations, and it is well suited for the rapid generation of large data sets. Presented in this paper are the basic principles that guide the experiment's design, the derivation of an underlying mathematical model that serves to interpret the data, and experimental results that confirm the model's predictive power.

## I. INTRODUCTION

The phase behavior of biomembrane mixtures has long been a topic of active research. In recent years, particularly intense interest has focused on so-called "lipid rafts," a type of domain structure thought to form spontaneously by lateral phase separation in membranes that are rich in cholesterol and certain sphingolipids [1,2]. Since 2001, a number of groups have published ternary phase diagrams for a variety of raft-like mixtures, based at least in part on experiments using confocal fluorescence microscopy (CFM) and giant unilamellar vesicles (GUVs) [e.g., 3-5].

CFM studies exploit the general tendency of fluorescent membrane probes to partition preferentially between coexisting membrane phases—labeling one phase more brightly than another—and they can indeed produce striking images of coexisting membrane phases [6]. By carefully mapping all the compositions that manifest phase separation, CFM experiments have proven a valuable tool for determining phase boundaries in raft-like membrane mixtures. However, for several technical reasons, CFM experiments have not proven capable of determining tie-line trajectories, so alternative techniques have been sought for this purpose [7]. Moreover, each of the protocols currently used to produce GUVs—a vesicular form required by CFM studies—passes the membrane mixture through an intermediary solvent-free solid state, a treatment that may give rise to artifactual phase separation [8].

This paper describes an experimental technique—termed "Steady-State Probe-Partitioning FRET," or simply SP-FRET—which has been developed specifically for the analysis of phase behavior in multi-component membrane mixtures. Like CFM, SP-FRET exploits the general tendency of fluorescent membrane probes to partition preferentially between coexisting membrane phases (Figure 1). However, SP-FRET experiments are cuvette-based and can therefore employ ordinary, polydisperse vesicle suspensions, without the need for GUVs or any other sort of specially prepared vesicles. Like CFM, SP-FRET can detect the presence or absence of phase domains, but SP-FRET requires no equipment more sophisticated than an ordinary steady-state fluorometer. And unlike CFM images, SP-FRET data can easily be interpreted via a simple model to yield not just phase boundaries, but also probe partition coefficients and tie lines in any phase-separating membrane mixture.

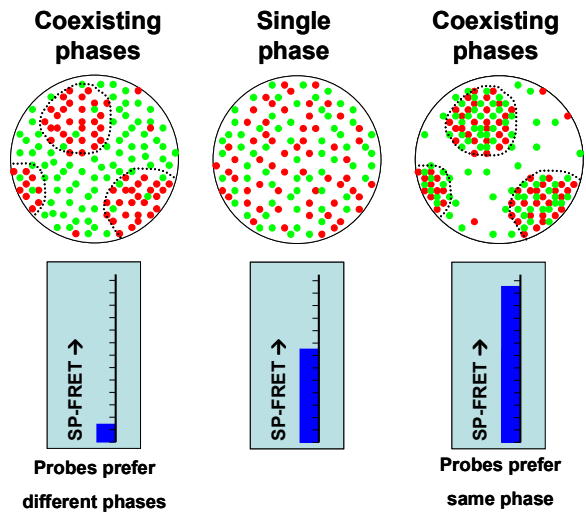

**FIG 1. (Color online) Probe-Partitioning Fluorescence Resonance Energy Transfer.** Membranes are labeled with trace quantities of fluorescent donor (green) and acceptor (red) probes. In a single-phase membrane, an intermediate FRET signal intensity is observed. In the presence of coexisting phases, however, the probes can partition preferentially between alternative environments, causing their local concentrations to rise or fall. If the probes prefer different phases, they are effectively separated, decreasing FRET. If the probes prefer the same phase, they are effectively clustered and the FRET signal increases.

The three main purposes of this paper are to (a) explain the basic principles that underlie SP-FRET experiments; (b) derive the mathematical model that serves to interpret SP-FRET data; and (c) present experimental results that demonstrate SP-FRET analysis in the context of a phase-separating membrane mixture. The results presented here have been chosen to demonstrate SP-FRET within the simplest possible system: a binary mixture of phospholipids manifesting coexisting fluid and solid membrane phases near room temperature. However,

[1] e-mail: jbuboltz@colgate.edu

more ambitious applications of SP-FRET (e.g., tie-line determination in ternary or quaternary mixtures) will be addressed in future papers.

## II. Modeling SP-FRET in a regime of coexisting phases

### A. Choice of FRET Efficiency Metric

Although a variety of FRET metrics are in active use [9], the most commonly used metric is the transfer efficiency, or the fractional decrease in donor fluorescence due to acceptor quenching, generally symbolized by the letter $E$,

$$E = 1 - \frac{F'_D}{F_D} = 1 - \frac{\tau'_D}{\tau_D}$$

where $F_D$ is the intensity of donor fluorescence in the absence of acceptor, $F'_D$ is donor fluorescence in the presence of acceptor, and $\tau_D, \tau'_D$ are the associated excited-state lifetimes [10]. While $E$ is conveniently dimensionless and has a natural range between zero and one, it does unfortunately require two measurements: donor fluorescence in both the presence and absence of acceptor. This means that two independent samples must be prepared and measured in order to assess $E$ at any given membrane composition.

In order to facilitate high resolution experiments over wide-ranging composition spaces, SP-FRET employs an alternative metric: $F_A^{Dex}$, the intensity of acceptor fluorescence under donor excitation. Although $F_A^{Dex}$ must be expressed in the arbitrary units of fluorescence intensity, this metric makes it possible to assess FRET efficiency without multiple samples. Although $F_A^{Dex}$ does not range between zero and one, it does range between zero and a maximum asymptotic limit. And as long as all relevant experimental parameters (e.g., spectrophotometer settings and global probe concentrations) are kept fixed throughout a given experiment, variations in $F_A^{Dex}$ can still be interpreted quantitatively in terms of the mathematical model that follows.

### B. Overview of SP-FRET Model

In order to model $F_A^{Dex}$ variation in a regime of $m$ coexisting membrane phases, it may first be noted that (i) on the timescale of typical excited state lifetimes ($\leq 10^{-8}$ s) there is little diffusive motion of membrane probes [11]; and that (ii) the characteristic spatial scale, $R_d$, of many membrane phase domains is much larger than the donor-acceptor Förster distance, $R_0$, of even the most efficient energy transfer probe pairs (i.e., $R_0 \sim 5$ nm). For both these reasons, the fraction of overall energy transfer that occurs across domain boundaries should be minimal[2] and the observed (i.e., global average) donor-excited acceptor fluorescence may be approximated as the sum of individual contributions $F_i \equiv \left( F_A^{Dex} \right)_i$ from within

each of the sample's coexisting phases:

$$F_A^{Dex} = \sum_{i=1}^{m} F_i \qquad (1)$$

The contribution from each phase can be further expressed as

$$F_i = S_i \cdot f_i \qquad (2)$$

where $f_i$ is the phase-specific $F_A^{Dex}$ signal originating from phase $i$ and $S_i$ is the Lever Rule scaling factor that specifies the fraction of the system in state $i$ at equilibrium. These two parameters, $S_i$ and $f_i$, will now be discussed in more detail, beginning with the scaling factors.

### C. Phase Scaling Factors

For any sample with coexisting phases, scaling factors may be calculated, provided that both the sample's global composition and the coexisting phase compositions are provided as input. For any membrane component $j$, its global mole fraction $\chi_j^0$ must be the $S_i$-weighted sum of its local mole fractions $\chi_j^i$ in the coexisting phases:

$$\chi_j^0 = \sum_{i=1}^{m} \chi_j^i \cdot S_i \qquad (3)$$

Therefore, if one considers a mixture of $n$ different components distributed among $m$ coexisting phases, it follows that any sample of global composition

$$\chi_1^0 \quad \chi_2^0 \quad \cdots \quad \chi_n^0$$

that falls within a particular regime of coexisting phase compositions

$$\begin{matrix} \chi_1^1 & \chi_1^2 & \cdots & \chi_1^m \\ \chi_2^1 & \chi_2^2 & \cdots & \chi_2^m \\ \vdots & \vdots & & \vdots \\ \chi_n^1 & \chi_n^2 & \cdots & \chi_n^m \end{matrix}$$

must form its coexisting phases in proportion to a certain set of scaling factors

$$S_1 \quad S_2 \quad \cdots \quad S_m$$

such that

$$\begin{pmatrix} \chi_1^1 & \chi_1^2 & \cdots & \chi_1^m \\ \chi_2^1 & \chi_2^2 & \cdots & \chi_2^m \\ \vdots & \vdots & & \vdots \\ \chi_n^1 & \chi_n^2 & \cdots & \chi_n^m \end{pmatrix} \begin{pmatrix} S_1 \\ S_2 \\ \vdots \\ S_m \end{pmatrix} = \begin{pmatrix} \chi_1^0 \\ \chi_2^0 \\ \vdots \\ \chi_n^0 \end{pmatrix} \qquad (4)$$

Because the Gibbs Phase Rule stipulates $m \leq n$, Equation 4

---

[2] In order to allow for the study of "nanoscopic" membrane domains, which have $R_d$ on the order of 100 nm or smaller [12,13], SP-FRET experiments can simply employ less efficient donor-acceptor pairs (i.e., probe combinations with reduced spectral overlap) in order to ensure that $R_0 \ll R_d$.

can always be solved for the scaling factors $\{S_i\}$ that are implied by a sample's global composition $\{\chi_j^0\}$ together with a particular set of coexisting phase compositions $\{\chi_j^i\}$. The $\{\chi_j^i\}$ can be inferred from $\nabla F_A^{Dex}$ (see section III E).

### D. Phase-Specific FRET Efficiencies

In contrast to the scaling factors—which simply reflect mass balance—the $f_i$ contain rather more information, convolving photo-physical differences between phase environments (e.g., average probe dipole orientations) together with variations in local donor and acceptor concentrations that occur due to differential probe partitioning between phases. However, for the purposes of SP-FRET analysis, one may describe these effects using a relatively simple expression

$$f_i(\chi_D^i, \chi_A^i) = \frac{C_0^i \cdot \chi_D^i \cdot \chi_A^i}{1 + C_1^i \cdot \chi_A^i} \qquad (5)$$

in which the phase specific $F_A^{Dex}$ signal originating from phase $i$ ($f_i$) is an explicit function only of local donor and acceptor concentrations ($\chi_D^i$ and $\chi_A^i$), and all photo-physical effects are folded into two constants ($C_0^i, C_1^i$) that are specific to that phase. Equation 5 is similar to a previously published acceptor-dependence expression adopted by Zacharias, *et al.* [14] on essentially phenomenological grounds: proper behavior in the limits of very low and very high acceptor concentration. Indeed, Eq. 5 does behave as it should in its limits: it predicts asymptotic approach to a limiting FRET efficiency as $\chi_A^i$ gets very large, and it predicts that FRET will vanish as either $\chi_D^i$ or $\chi_A^i$ approach zero.

However, Eq. 5—which describes donor-excited acceptor fluorescence between freely diffusing populations of donors and acceptors—can actually be derived from a simple chemical kinetic model [15]. The conditions under which Eq. 5 is valid are those in which (i) probe self-quenching is negligible (i.e., dilute probe concentrations), and (ii) excited-state concentrations remain both low (i.e., moderate excitation intensity) and constant (i.e., steady-state fluorescence). The form of Eq. 5 has recently been validated experimentally by donor-acceptor titration experiments carried out in three dissimilar membrane-phase environments (L$_\alpha$, L$_\beta$, and L$_o$). These experiments will be described in a forthcoming paper [ibid].

In fact, SP-FRET experiments are best carried out at particularly low concentrations of acceptor (i.e., $\chi_A^i << 1/C_1^i$),

conditions under which Eq. 5 approaches linearity:

$$f_i(\chi_D^i, \chi_A^i) \approx C_0^i \cdot \chi_D^i \cdot \chi_A^i$$

(e.g., $\chi_A^0 \le 10^{-4}$ for alkylcarbocyanine probes [*ibid.*]). Under these conditions, Eq. 5 may be simplified even further, defining the one remaining photo-physical constant $C_0^i$ in terms of $f_i^0$, the experimentally observed $F_A^{Dex}$ at the phase boundary where the sample consists entirely of phase $i$:

$$f_i(\chi_D^i, \chi_A^i) \approx \left( \frac{f_i^0}{\chi_D^0 \cdot \chi_A^0} \right) \cdot \chi_D^i \cdot \chi_A^i \qquad (6)$$

Equation 6 is a simple expression that serves to describe local variations in $F_A^{Dex}$ among coexisting membrane phases under conditions of very dilute donor and acceptor concentrations. Photo-physical differences between phases are accounted for by the experimentally determined $f_i^0$ term, while the $\chi_D^i \cdot \chi_A^i$ term describes the dependence on local probe concentrations.

### E. Tie Line Trajectories and Partition Coefficients

Because Equation 6 specifies the manner in which $\left( F_A^{Dex} \right)_i$ depends on changes in $\chi_D^i$ and $\chi_A^i$, an explicit expression must now be sought that describes how these changes are expected to occur as the global-average $F_A^{Dex}$ is measured over a range of membrane compositions. Such an expression can easily be obtained if the description is constrained to lie along a tie line trajectory.

Anywhere along a tie line, the thermodynamic properties of all coexisting phases are invariant—the phases vary only in extent. Therefore, along a tie line one can define constants called partition coefficients that characterize the relative concentrations of any probe molecule $P$ partitioning between two coexisting phases, $i$ and $i+1$:

$$K_i^P \equiv \frac{\chi_P^{i+1}}{\chi_P^i}$$

Because each probe's global concentration $\chi_P^0$ is fixed, it follows that specifying all the partition coefficients $\{K_i^P\}$ for a given probe is equivalent to specifying the concentration of that probe within each of the phases. For a system with $m$ coexisting phases, Equation 3 becomes

$$\chi_P^0 = \chi_P^1 \cdot S_1 + \chi_P^1 K_1^P \cdot S_2 \cdots + (\chi_P^1 K_1^P K_2^P \cdots K_{m-1}^P) \cdot S_m$$

where the subscript serves to remind that the mixture component under consideration is a probe, present only in trace quantities. This expression can be recast as

$$\chi_P^1 = \frac{\chi_P^0}{\sum_{k=1}^{m} \left( S_k \prod_{l=0}^{k-1} K_l^P \right)}$$

where $K_0^P \equiv 1$, and each local probe concentration $\chi_P^i$ can therefore be computed from the following equation

$$\chi_p^i = \frac{\chi_P^0 \prod_{l=0}^{i-1} K_l^P}{\sum_{k=1}^{m} (S_k \prod_{l=0}^{k-1} K_l^P)} \qquad (7)$$

given a set of $m-1$ partition coefficients $\{K_i^P\}$ for that probe.

### F. General Expression

Combining equations 1, 2, 6 and 7 leads to a general expression for dilute-probe experiments that describes the experimentally observed SP-FRET signal at any point along a tie line traversing a regime of $m$ coexisting phases:

$$F_A^{Dex} = \sum_{i=1}^{m} S_i \cdot f_i^o \cdot \left( \frac{\prod_{l=0}^{i-1} K_l^D K_l^A}{\sum_{k=1}^{m} (S_k \prod_{l=0}^{k-1} K_l^D) \sum_{k=1}^{m} (S_k \prod_{l=0}^{k-1} K_l^A)} \right) \qquad (8)$$

Equation 8 contains $2(m-1)$ fitting parameters: the probe partition coefficients for donor and acceptor species. Given the coexisting compositions for any phase-separated membrane mixture—a condition equivalent to specifying phase boundaries and tie line trajectories—Equation 8 describes how the experimentally observed SP-FRET signal can be expected to vary throughout the regime of phase coexistence.

### III. MATERIALS AND METHODS

#### A. Chemicals

DLPC and DPPC were purchased from Avanti Polar Lipids and purity was confirmed by thin layer chromatography on washed, activated silica gel plates as previously described [16]. Dialkylcarbocyanine probes (i.e., DiO and DiI species) were from Molecular Probes and dehydroergosterol (DHE) was from Sigma-Aldrich Chemical Corp. PIPES buffer and disodium EDTA were purchased from Fluka Chemie AG. Aqueous buffer (2.5mM PIPES pH 7.0, 250mM KCl, 1mM EDTA) was prepared from 18 MΩ water (Barnstead E-Pure) and filtered through a 0.2 μm filter before use.

#### B. Donor and Acceptor Probes

FRET probes (i.e., donor and acceptor pair combinations) were chosen for their favorable spectral overlap, as well as their tendency to show pronounced preference for either the $L_\alpha$ (i.e., fluid) or the $L_\beta$ (i.e., solid) membrane phase. DiO and DiI carbocyanine dyes are commercially available with 18-carbon chains that can be either fully saturated (18:0) or doubly cis-unsaturated (18:2). The 18:0 species prefer $L_\beta$ over $L_\alpha$, whereas the 18:2 species prefer $L_\alpha$ over $L_\beta$. DHE, the cholesterol analog, was expected to prefer a disordered $L_\alpha$ environment over the more ordered $L_\beta$ environment.

#### C. Sample Preparation

Specified sample compositions ($1.0 \times 10^{-7}$ moles total lipid per sample) were prepared in 13 x 100 mm screw cap tubes by combining appropriate volumes of chloroform-based lipid and probe stock solutions using gastight Hamilton volumetric syringes. 1.0 ml of aqueous buffer was then added to each tube, and the chloroform was removed by a modified version of the Rapid Solvent Exchange procedure [8]. Samples were sealed under argon, placed in a temperature controlled water bath at 45.0°C, and then slowly cooled (~ 4°C/hour) to 20.0°C where they were held for two days before measurement. Probe/lipid ratios were fixed at 1/10,000 for the carbocyanines and 1/500 for DHE.

#### D. Fluorescence Measurements

Fluorescence measurements were carried out on a Hitachi F4500 fluorescence spectrophotometer in photometry mode (10.0 sec integration; 5.0/10.0 mm slits) using a temperature-controlled cuvette holder (Quantum Northwest, Inc). For $F_A^{Dex}$ measurements, excitation/emission channels were set to either 325/505nm (DHE→DiO) or 430/570nm (DiO→DiI). Accurate spectral deconvolution is essential for SP-FRET experiments, so rigorous background, bleed-through and 'cross-talk' corrections [9] were provided for. In brief, the F4500 was set up to record four channel combinations for each sample: a scattering signal (430/430nm) and three separate fluorescence signals ($F_D^{Dex}, F_A^{Dex}, F_A^{Aex}$). Calibration standards (i.e., probe-free and single-probe samples) were included in every set of measurements, and periodic closed-shutter integrations were collected for dark current correction. After the raw fluorescence data had been corrected for each possible form of background signal (i.e., dark current, scattering and spurious fluorescence), spectral deconvolution was performed, with the calibration standards serving as quality control samples.

#### E. SP-FRET Profile Analysis

Experimentally determined DLPC/DPPC SP-FRET profiles (i.e., $F_A^{Dex}$ vs. $\chi_{DPPC}$) were fit to Eq. 9 (see Results) by two different techniques: Exhaustive exploration of $K^D, K^A$ space and Newton-Raphson optimization starting from randomly initialized $K^P$'s. Both strategies entail concomitant fitting of two or more profiles in order to resolve the fit-degeneracy inherent in Eq. 9 (see Results section). In these mixtures, probe partition coefficients were defined as $K^P \equiv \frac{\chi_P^{solid}}{\chi_P^{fluid}}$.

Exhaustive fits were carried out on conjugate profiles—two SP-FRET data sets sharing one probe in common, both collected at the same temperature along the same trajectory in composition space—using a custom fitting routine written in Java. The routine searches 100 points along each axis of a 3-dimensional fitting space (three independent $K^P$'s), seeking to minimize a reduced chi-squared parameter ($\chi_{red}^2$) that convolves the goodness of fit for both profiles. Optimization fits were carried out on a four-profile set comprising three conjugate pairs (to smooth the $\chi_{red}^2$ gradient and deepen the best-fit minima) using commercially available software (Systat 11, Systat Software, Inc).

Phase boundaries were identified as compositions at which $\left|\left\langle \nabla F_A^{Dex} \right\rangle\right|$ was maximized. Differences between global and local probe concentrations are maximized just inside a phase boundary—where the fraction of the minor phase approaches zero—so phase boundaries are places where the observed SP-FRET signal can be expected to change most rapidly. $L_\alpha$-$L_\beta$ phase boundaries were therefore assigned to the two DLPC/DPPC compositions at which the average $dF_A^{Dex}/d\chi_{DPPC}$ of all four experimentally determined SP-FRET profiles was greatest. Excluding $\chi_{DPPC} > 0.92$ (due to pronounced photo-physical effects as $\chi_{DPPC} \rightarrow 1$), these compositions were identified as $\chi_{DPPC}^{fluidus} = 0.235$ and $\chi_{DPPC}^{solidus} = 0.785$ in DLPC/DPPC at 20°C, values consistent with previously published results [17,18].

## IV. EXPERIMENTAL RESULTS

A binary mixture is the simplest possible example of a system that can phase separate. In such a system, $n = m = 2$; and the composition space is one-dimensional, so that any coexistence regime must necessarily constitute a tie line. In this case, Equation 8 reduces to a simple function of just one independent variable, $S_2$, with two fitting parameters, $K^D$ and $K^A$:

$$F_A^{Dex} = \frac{f_1^o + S_2(f_2^o K^D K^A - f_1^o)}{[1 + (K^D - 1)S_2][1 + (K^A - 1)S_2]} \qquad (9)$$

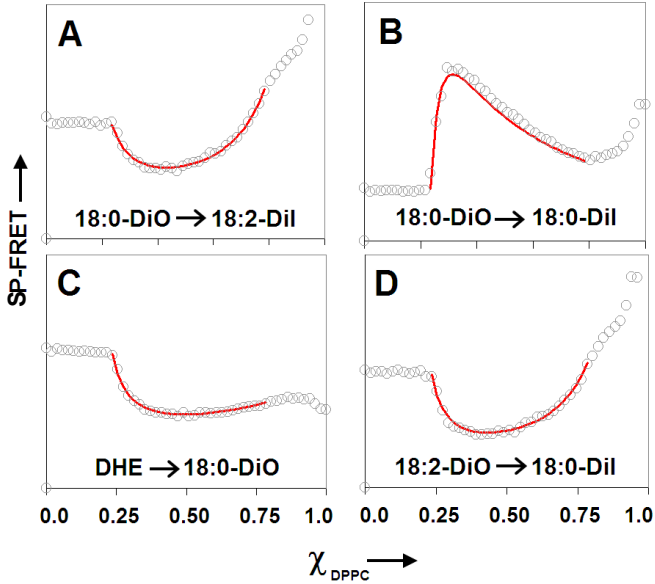

**FIG 2.** (Color online) **SP-FRET profiles for four different combinations of donor → acceptor probes, carried out in DLPC/DPPC at 20.0°C.** SP-FRET efficiency ($F_A^{Dex}$, arbitrary units) is plotted vs. mole fraction of DPPC. In each profile, a distinct regime of reduced efficiency or enhanced efficiency corresponds to coexisting $L_\alpha$ (fluid) and $L_\beta$ (solid) phases between $\chi_{DPPC}^{fluidus} = 0.235$ and $\chi_{DPPC}^{solidus} = 0.785$. Reduced efficiency is evident in panels A, C and D, in which donors and acceptors prefer different phases. In panel B, enhanced efficiency indicates the preference of both probes for the DPPC-rich $L_\beta$ phase. Red lines correspond to best-fit curves in accordance with Equation 9 and the following partition coefficients: $K^{18:0-DiO} = 7.4 \pm 0.2$; $K^{18:2-DiO} = 0.23 \pm 0.1$; $K^{18:0-DiI} = 7.5 \pm 0.3$; $K^{18:2-DiI} = 0.22 \pm 0.01$; $K^{DHE} = 0.61 \pm 0.02$.

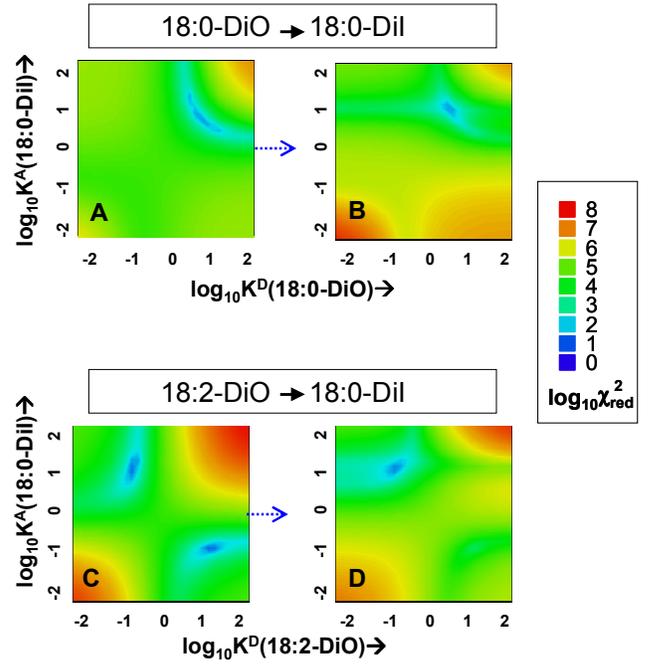

**FIG 3.** (Color online) **Resolution of fit degeneracy by coupled fitting of conjugate profiles.** Goodness of fit ($\log_{10}\chi_{red}^2$) data is plotted over a 10,000-fold range of both donor and acceptor $K^P$ values for two of the SP-FRET profiles from Fig. 2. Left-hand plots illustrate the degeneracy observed when either of the plots is fit in isolation to Eq. 9: The 18:0-DiO→18:0-DiI profile is fit well (i.e., $\chi_{red}^2 \approx 1$) by a range of complementary $K^D, K^A$ values (blue arc, panel A) and the 18:2-DiO→18:0-DiI profile is fit by two symmetric loci in $K^D, K^A$ space (blue spots, panel C). Right-hand plots show that the degeneracy is resolved when the fits are coupled (i.e., using a single merged $\chi_{red}^2$ parameter), in which case both SP-FRET curves are best fit by unique $K^D, K^A$ combinations (panels B and D).

Figure 2 shows four experimentally determined SP-FRET profiles generated in a phase-separating binary mixture and fit according to Eq. 9. Four independent series of DLPC/DPPC suspensions were prepared at 20.0°C and to each was added a different combination of donor and acceptor probes. In three of the experiments (Fig. 2A,C,D), donors and acceptors prefer opposite phases, creating a regime of reduced efficiency within each profile. In the fourth experiment (Fig. 2b), both donors and acceptors partition preferentially into the same phase, creating a regime of enhanced efficiency. It should be emphasized that all the data in Fig. 2—four different profiles from four separate experiments—were fit with a total of just five free parameters: the probe partition coefficients.

It must also be noted that the five best-fit values reported in the Fig. 2 legend were not derived from isolated fits of individual SP-FRET profiles. Equation 9 is degenerate with respect to $K^D$ and $K^A$, so unique best-fit $K^P$'s cannot be obtained by fitting Eq. 9 to a single profile.

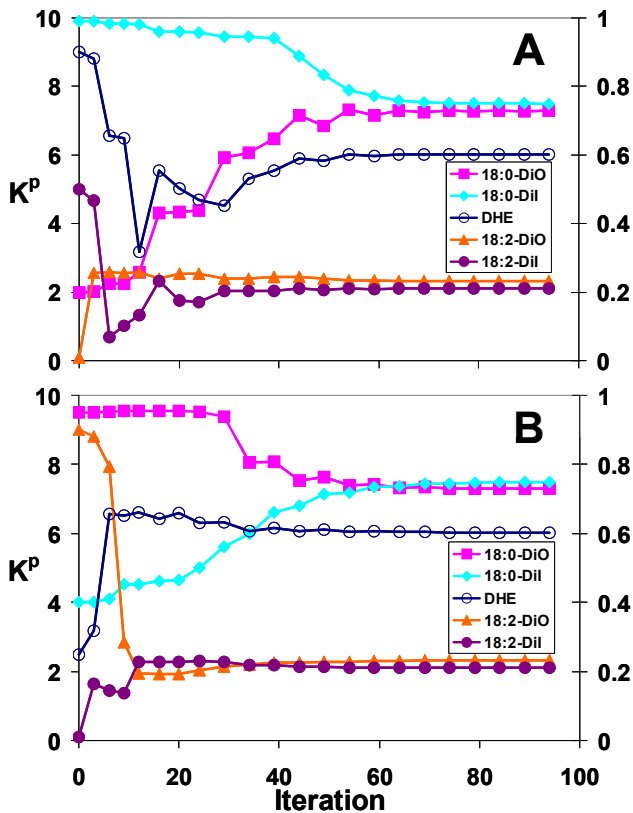

**FIG 4. (Color online) An optimization approach yields the same best-fit $K^P$ values.** In the experiments shown above, Newton's method was used for the simultaneous optimization of all four curve fits shown in Fig. 2. Panels A and B show two independent trials starting from different initialization conditions. The left vertical axis represents $K^P$ values for the two 18:0 probe species, while the right axis shows $K^P$ values for both the 18:2 probes and DHE. All five $K^P$ values converge to the same values reported in the legend for Fig. 2.

However, coupling the fits of two conjugate profiles (*i.e.*, profiles that share one probe in common) resolves this degeneracy, as shown in Fig. 3. Color-coded $\chi^2_{red}$ plots in $K^D, K^A$ space are shown for two SP-FRET experiments: the profile in Fig. 2b (Fig. 3 upper panels), and the profile in Fig 2d (Fig. 3 lower panels). The left-hand $\chi^2_{red}$ plots correspond to fits of each profile in isolation and illustrate the fit degeneracy. The right-hand plots, however, were produced by coupling the fits of the two conjugate SP-FRET profiles, so that the degeneracy is resolved and unique best-fit combinations of $K^D$ and $K^A$ are obtained. Figure 4 shows that Newton-Raphson optimization applied simultaneously to all four fits in Fig. 2 produces the same $K^P$ values as the exhaustive conjugate-profile fits illustrated in Fig. 3.

## V. DISCUSSION

The SP-FRET results presented above are both self-consistent and in agreement with previously published studies of DLPC/DPPC phase behavior near room temperature [4,17,18]. Although all five partition coefficients were treated as completely independent fitting parameters, both the 18:0-DiO and 18:0-DiI $K^P$'s converged to the same $L_\beta$-preferring value (~7.4), while both the 18:2-DiO and 18:2-DiI $K^P$'s converged to the same $L_\alpha$-preferring value (~0.22). DHE, the only probe expected to show unique partitioning behavior, did in fact yield a unique best-fit partition coefficient: $K^{DHE}$ ~0.60. When parameterized with these five best fit $K^P$ values, Eq. 8 is in very good agreement ($\chi^2_{red} \approx 1$ without any structure in the residuals) with all four of the experimentally determined data sets in Fig. 2.

An essential design aspect of SP-FRET experiments is the careful consideration of the many different possible combinations of donor and acceptor probes. There are three main criteria to consider when choosing suitable probe combinations, each of which will now be discussed briefly.

First, *all SP-FRET probes must have sufficiently strong partitioning behavior*. In other words, it is important to choose probes that will manifest $K^P$ values as different from unity as possible, in order to maximize the sensitivity of $F_A^{Dex}$ to domain formation near the phase boundaries. As an illustration, compare the fluidus and solidus boundaries in Fig. 2c. The fluidus location is clearly marked by a rapid change in $F_A^{Dex}$ caused by strong (7.4-fold) partitioning of the solid-preferring acceptor into the minority $L_\beta$-phase domains. In contrast, the solidus location is obscured by relatively weak (1.6-fold) partitioning of the fluid-preferring donor into the minority $L_\alpha$-phase domains.

Second, *the set of probes employed should manifest complementary partitioning*. In other words, if three different probes are used together in each sample—in order to form conjugate probe pairs, for example—then it should never be the case that all three probes partition into the same phase. Rather, it should be arranged that each of the coexisting membrane phases will be preferred by at least one of the probes employed, so that the location of every phase boundary is marked by changes in $F_A^{Dex}$ caused by probe partitioning into minority domains near that boundary. Fig. 2b serves to illustrate the importance of this particular probe-choice criterion: because neither of the 18:0 probes partitions into the minority $L_\alpha$-domains near the solidus, this phase boundary is more difficult to identify.

Third, *each donor-acceptor pair should have an $R_0$ value significantly smaller than $R_d$*. For example, $L_\alpha$-$L_\beta$ coexistence (as in the current work) is amenable to probe pairs with quite large Förster distances (i.e., $R_0 \sim 50$ Å), but pairs with less spectral overlap (e.g., $R_0 \leq 5$ Å) should be chosen for the study of nanoscopic membrane domains [12,13].

The SP-FRET technique described in this paper is certainly not the first FRET-based strategy to be applied to the study of membrane phase behavior. It has long been recognized that energy transfer between fluorescent membrane probes is eminently suited to the detection and study of coexisting membrane phase domains [19], and a number of groups have adapted FRET experiments to this end during the last decade, or so [e.g., 4,20-22]. Indeed, Loura, de Almeida, Fedorov and Prieto have published several excellent studies [13, 23-27] based on time-resolved measurements of transfer efficiency, $E(t)$, and Silvius and Nabi have recently produced a comprehensive review of FRET-based studies of membrane microdomains [28].

The advantage of SP-FRET experiments is that they are comparatively simple to perform and can be easily adapted for the high-resolution mapping of phase behavior over a wide-ranging composition space. In order to map SP-FRET using a single sample at each membrane composition, conjugate probe pairs can simply be included in each sample (e.g., DHE + 18:2-DiO + 18:2-DiI) so that conjugate $F_A^{Don}$ profiles are generated from a single sample set. And since SP-FRET measurements can be carried out rapidly on an ordinary steady-state fluorometer, it is possible to generate large data sets with relative ease. As long as suitable probe combinations are chosen, SP-FRET should prove to be a robust and easily implemented tool for the general study of composition-dependent membrane phase behavior.

ACKNOWLEDGMENTS

This work was supported by Research Corporation Award CC6814.